\documentclass[dvips]{acta}
\usepackage{supertabular,lscape,epsfig}
\usepackage{lscape,epsfig}
\usepackage{amssymb}
\usepackage{amsmath}
\usepackage{float}
\mathchardef\mhyphen="2D
\usepackage{placeins}
\DeclareSymbolFont{ppa}{OT1}{ppl}{m}{it}
\DeclareMathSymbol{\vv}{\mathalpha}{ppa}{'166}

\SetPages{0}{0}

\SetVol{69}{2019}

\begin{document}
\newcommand\pvalue{\mathop{p\mhyphen {\rm value}}}
\newcommand{\TabApp}[2]{\begin{center}\parbox[t]{#1}{\centerline{
  {\bf Appendix}}
  \vskip2mm
  \centerline{\small {\spaceskip 2pt plus 1pt minus 1pt T a b l e}
  \refstepcounter{table}\thetable}
  \vskip2mm
  \centerline{\footnotesize #2}}
  \vskip3mm
\end{center}}

\newcommand{\TabCapp}[2]{\begin{center}\parbox[t]{#1}{\centerline{
  \small {\spaceskip 2pt plus 1pt minus 1pt T a b l e}
  \refstepcounter{table}\thetable}
  \vskip2mm
  \centerline{\footnotesize #2}}
  \vskip3mm
\end{center}}

\newcommand{\TTabCap}[3]{\begin{center}\parbox[t]{#1}{\centerline{
  \small {\spaceskip 2pt plus 1pt minus 1pt T a b l e}
  \refstepcounter{table}\thetable}
  \vskip2mm
  \centerline{\footnotesize #2}
  \centerline{\footnotesize #3}}
  \vskip1mm
\end{center}}

\newcommand{\MakeTableH}[4]{\begin{table}[H]\TabCap{#2}{#3}
  \begin{center} \TableFont \begin{tabular}{#1} #4 
  \end{tabular}\end{center}\end{table}}

\newcommand{\MakeTableApp}[4]{\begin{table}[p]\TabApp{#2}{#3}
  \begin{center} \TableFont \begin{tabular}{#1} #4 
  \end{tabular}\end{center}\end{table}}

\newcommand{\MakeTableSepp}[4]{\begin{table}[p]\TabCapp{#2}{#3}
  \begin{center} \TableFont \begin{tabular}{#1} #4 
  \end{tabular}\end{center}\end{table}}

\newcommand{\MakeTableee}[4]{\begin{table}[htb]\TabCapp{#2}{#3}
  \begin{center} \TableFont \begin{tabular}{#1} #4
  \end{tabular}\end{center}\end{table}}

\newcommand{\MakeTablee}[5]{\begin{table}[htb]\TTabCap{#2}{#3}{#4}
  \begin{center} \TableFont \begin{tabular}{#1} #5 
  \end{tabular}\end{center}\end{table}}

\newcommand{\MakeTableHH}[4]{\begin{table}[H]\TabCapp{#2}{#3}
  \begin{center} \TableFont \begin{tabular}{#1} #4 
  \end{tabular}\end{center}\end{table}}

\newfont{\bb}{ptmbi8t at 12pt}
\newfont{\bbb}{cmbxti10}
\newfont{\bbbb}{cmbxti10 at 9pt}
\newcommand{\uprule}{\rule{0pt}{2.5ex}}
\newcommand{\douprule}{\rule[-2ex]{0pt}{4.5ex}}
\newcommand{\dorule}{\rule[-2ex]{0pt}{2ex}}
\def\thefootnote{\fnsymbol{footnote}}
\begin{Titlepage}
\Title{Final Release of the OGLE Collection of Cepheids and RR~Lyrae Stars in the Magellanic System. The Outer Regions\footnote{Based on observations
obtained with the 1.3-m Warsaw telescope at the Las Campanas Observatory of
the Carnegie Institution for Science.}}
\Author{I.~~S~o~s~z~y~ñ~s~k~i$^1$,~~
A.~~U~d~a~l~s~k~i$^1$,~~
M.\,K.~~S~z~y~m~a~ñ~s~k~i$^1$,~~
P.~~P~i~e~t~r~u~k~o~w~i~c~z$^1$,\\
J.~~S~k~o~w~r~o~n$^1$,~~
D.~~S~k~o~w~r~o~n$^1$,~~
R.~~P~o~l~e~s~k~i$^2$,~~
S.~~K~o~z~³~o~w~s~k~i$^1$,\\
P.~~M~r~ó~z$^1$,~~
K.~~U~l~a~c~z~y~k$^3$,~~
K.~~R~y~b~i~c~k~i$^1$,~~
P.~~I~w~a~n~e~k$^1$~~
and~~M.~~W~r~o~n~a$^1$
}
{$^1$Warsaw University Observatory, Al.~Ujazdowskie~4, 00-478~Warszawa, Poland\\
e-mail: soszynsk@astrouw.edu.pl\\
$^2$Department of Astronomy, Ohio State University, 140 W. 18th Ave., Columbus, OH~43210, USA\\
$^3$Department of Physics, University of Warwick, Gibbet Hill Road, Coventry, CV4~7AL,~UK}
\Received{June 21, 2019}
\end{Titlepage}

\Abstract{We present the final release of the OGLE collection of classical
  pulsators (Cepheids and RR~Lyr stars) in the Large and Small Magellanic
  Clouds. The sky coverage has been increased from 670 to 765 square
  degrees compared to the previous edition of the collection. We also add
  some Cepheids and RR~Lyr stars found by the Gaia team and reclassify
  three Cepheids. Ultimately, our collection consists of 9650 classical
  Cepheids, 343 type~II Cepheids, 278 anomalous Cepheids, and 47\,828
  RR~Lyr stars inside and toward the Magellanic System.}{Stars:
  variables: Cepheids -- Stars: variables: RR~Lyrae -- Stars: oscillations
  (including pulsations) -- Magellanic Clouds -- Catalogs}

\Section{Introduction}
Cepheids and RR~Lyr stars (collectively referred to as classical
pulsators) play a key role in determining the cosmic distance scale,
tracing young (classical Cephe\-ids) and old (RR~Lyr stars) stellar
populations, and testing stellar models. The Optical Gravitational Lensing
Experiment (OGLE) has discovered an unprecedentedly large sample of
classical pulsators in the Magellanic System. The OGLE-II catalogs of
Cepheids in the Magellanic Clouds were published 20~years ago (\eg Udalski
\etal 1999ab). At that time, the OGLE fields covered only central regions
of both galaxies, less than 7 square degrees in total. The survey footprint
was increased to 54 square degrees in the OGLE-III catalogs of Cepheids and
RR~Lyr stars in the Large Magellanic Cloud (LMC) and Small Magellanic Cloud
(SMC, \eg Soszyñski \etal 2008, 2009, 2010). The most recent edition of the
OGLE Collection of Variable Stars (OCVS, \eg Soszyñski \etal 2017, 2018)
contains over 10\,000 classical, type~II, and anomalous Cepheids and over
46\,000 RR~Lyr variables in the area of 670 square degrees covering both
Magellanic Clouds and the Magellanic Bridge.

These nearly complete samples of classical pulsators in the Magellanic
System have been subject to extensive theoretical and empirical
investigations. The Cepheid period--luminosity (PL) and period--Wesenheit
index (PW) relations, their linearity and metallicity sensitivity in
various passbands have been studied by Gar\-c{\'{\i}}a-Varela \etal (2016),
Groenewegen and Jurkovic (2017b), Wielgórski \etal (2017), and Gieren \etal
(2018), mentioning only the most recent publications. The PL and PW
relations for classical pulsators have been used to measure distances to
the Magellanic Clouds (\eg Das \etal 2018, Gieren \etal 2018) and to
explore the three-dimensional structure of these galaxies (\eg Inno \etal
2016, Jacyszyn-Dobrzeniecka \etal 2016, 2017, 2019, Ripepi \etal 2017, Deb
\etal 2018, Muraveva \etal 2018, Prudil \etal 2018). The OGLE light curves
have been examined in detail to reveal physical properties of Cepheids and
RR~Lyr stars (\eg Skowron \etal 2016, Smolec and ¦niegowska 2016, Bhardwaj
\etal 2017, Marconi \etal 2017, Smolec 2017, Das \etal 2018, S{\"u}veges
and Anderson 2018ab). The evolutionary status of type~II and anomalous
Cepheids in the Magellanic Clouds was discussed by Gautschy and Saio
(2017), Groenewegen and Jurkovic (2017a), and Iwanek \etal (2018). Among 75
million stars observed by OGLE in the Magellanic System, we discovered new
subtypes of classical pulsators: peculiar W~Vir stars (Soszyñski \etal
2008), anomalous double-mode RR~Lyr stars (Soszyñski \etal 2016b),
first-overtone type~II Cepheids (Soszyñski \etal 2019), and a number of
Cepheids in eclipsing binary systems (Udalski \etal 2015b, Pilecki \etal
2018 and references therein).

In this paper, we extend the OGLE sky coverage of the Magellanic System to
about 765 square degrees by including fields in the outskirts of the LMC
and regions to the south of the Magellanic Bridge. We find about ten new
Cepheids and over 1100 RR~Lyr stars in this extended area. We also
reclassify a few previously published Cepheids and we add 232 classical
pulsators (mostly RR~Lyr variables) included in the Gaia DR2 catalog
(Clementini \etal 2019) and located in the previously analyzed region.

This paper is structured as follows. In Section~2, we discuss the OGLE
observations and data reduction. Section~3 describes methods used to select
and classify classical pulsators. In Section~4, we present our collection
itself and its cross-correlation with the International Variable Star Index
and Gaia DR2 Catalog of Cepheids and RR~Lyr stars. We discuss and summarize
our results in Sections~5 and~6.

\Section{Observations and Data Reduction}
\vspace*{-5pt}
Our study uses the {\it I}- and {\it V}-band photometric time-series data
collected during the OGLE-IV project (Udalski \etal 2015a). Observations
were conducted using the 1.3-m Warsaw telescope at Las Campanas Observatory
(operated by the Carnegie Institution for Science) in Chile. The telescope
is equipped with a mosaic camera composed of 32 CCDs, each with 2048 by
4096 pixels, providing a field of view of 1.4 square degrees on the sky.

The area in the outskirts of the LMC (56 fields located from 6 to 13
degrees from the LMC center) and regions in the southern ends of the
Magellanic System (27 fields with declination
$-83\arcd\lesssim\delta\lesssim-80\arcd$), hereafter outer fields, were
observed between August 2017 and April 2019. From about 100 to 150 epochs
per star in the Cousins {\it I}-band and from 10 to 25 observations in the
Johnson {\it V}-band have been secured during that period. The total OGLE
sky coverage of the Magellanic System has increased to 765 square
degrees. Detailed description of the instrumentation, photometric
reductions and astrometric calibrations of the OGLE data is provided in
Udalski \etal (2015a).

\vspace*{-7pt}
\Section{Identification and Classification of Classical Pulsators}
\vspace*{-9pt}
Compared to the previous editions of the OGLE collection of Cepheids and
RR~Lyr stars in the Magellanic Clouds (\eg Soszyñski \etal 2017, 2018), we
made some improvements to the procedure of the variable stars selection. As
before, we performed a period search for all sources observed in the {\it
I}-band in the extended OGLE fields. We used the {\sc Fnpeaks}
code\footnote{\it http://helas.astro.uni.wroc.pl/deliverables.php?lang=en\&active=fnpeaks}
by Z.~Ko³aczkowski.

In the first stage of our variability selection and classification process,
we visually inspected light curves with the strongest periodic
signal. Through this process non-variable and artificial objects were
filtered out from our sample. Objects that passed our criteria were tagged
as pulsators, eclipsing binaries and other variable stars.

\begin{figure}[p]
\hglue-3mm
{\includegraphics[width=13.1cm]{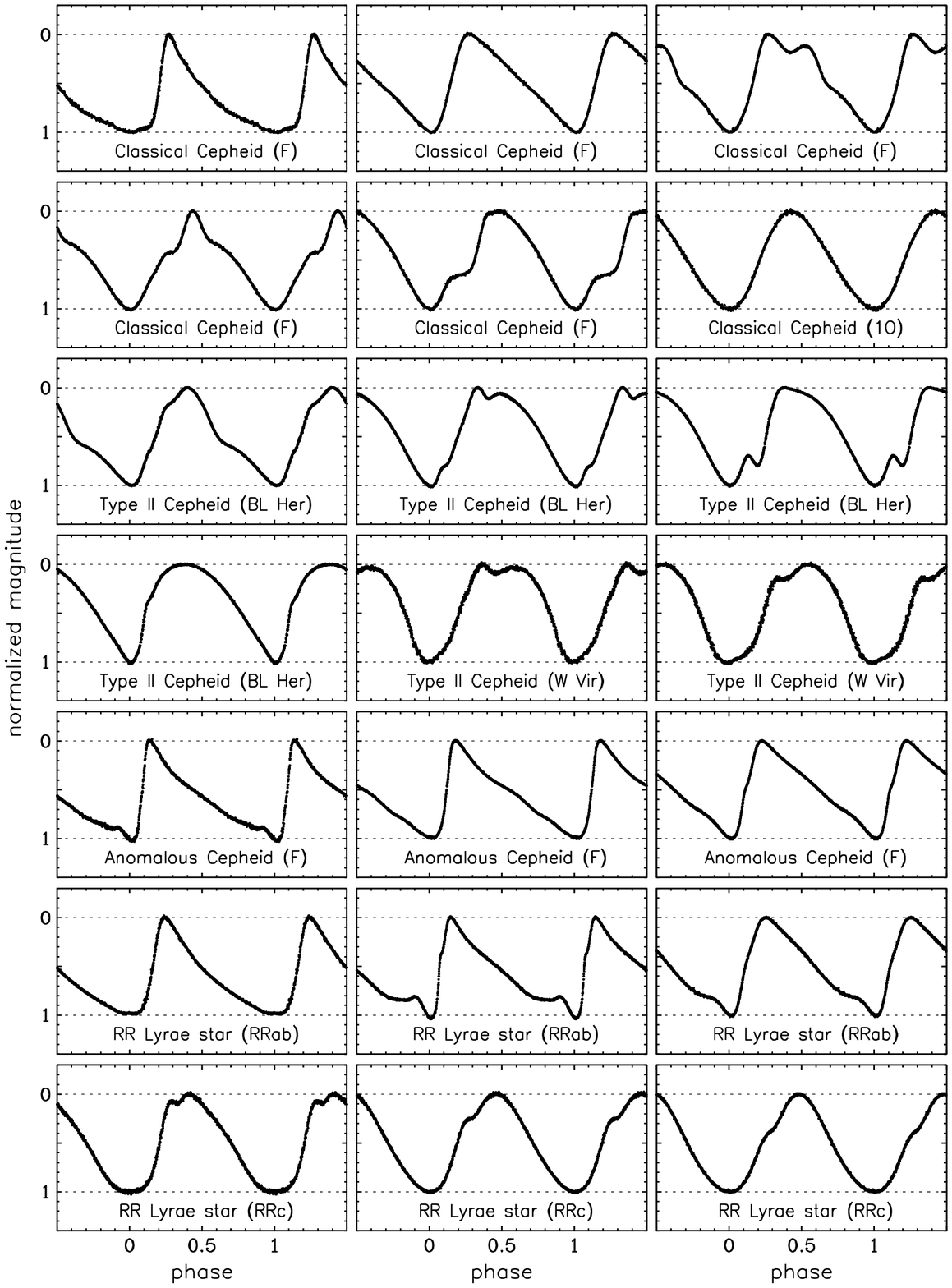}} 
\vspace*{2pt}
\FigCap{Examples of template light curves of classical pulsators. The
templates were obtained from the {\it I}-band light curves of bright,
well-sampled variables detected in the Galactic bulge and LMC. Types and
subtypes of individual classical pulsators are provided in panels.}
\end{figure}

In the second stage, we applied a template fitting method to all other
light curves. The algorithm was the same as we adopted to select over
450\,000 eclipsing and ellipsoidal binary systems in the Galactic bulge
(Soszyñski \etal 2016c), but in this case we used template light curves of
classical pulsators. These templates were obtained from 118 {\it I}-band
light curves of bright, single-mode classical, type~II, anomalous Cepheids,
and RR~Lyr stars observed by OGLE in the Galactic bulge and LMC. We
selected the best sampled light curves representing a wide range of
shapes. The Julian Dates of the individual observations were transformed to
the pulsation phases and averaged in 1000 bins. The magnitudes were
normalized in such a way that the maximum brightness of every template was
zero and the amplitude was equal to~1. Examples of our template light
curves are shown in Fig.~1.

The template fitting was carried out for all stars observed in the outer
regions of the Magellanic System, but also for the photometry collected in
previously analyzed OGLE-IV fields in the Magellanic Systems. We visually
verified each star automatically classified through this method. This
approach allowed us to improve the completeness of the OGLE collection of
Cepheids and RR~Lyr stars.

The vast majority of the identified classical pulsators turned out to be
RR~Lyr variables. Our classification primarily relies on the light curve
morphology quantitatively described by the Fourier coefficients $\phi_{21}$
and $\phi_{31}$ plotted against the logarithm of the period. The secondary
classification criterion was based on the position of a star on the PL and
PW diagrams, and, in the case of double-mode pulsators, on the position in
the period \vs period ratio (Petersen) diagram. As a result, we found
1157 RR~Lyr stars, two classical Cepheids, six anomalous Cepheids, and
three type~II Cepheids in the outer fields of the Magellanic
System. Additionally, the light curve template fitting and
cross-identification with the Gaia DR2 catalog of classical pulsators
(Clementini \etal 2019, see Section~4) increased the number of RR~Lyr
variables in the central regions of the Magellanic Clouds by 229, classical
Cepheids by two, and anomalous Cepheids by one object. We also reclassified
three variables located in the Magellanic Bridge from classical Cepheids
into anomalous Cepheids (one of them located in the Milky Way halo). The
old and new identifiers of these objects are listed in Table~1. The new and
reclassified variables have already been used in the studies of the
Magellanic Bridge by Jacyszyn-Dobrzeniecka \etal (2019).

\MakeTableee{
l@{\hspace{12pt}} l@{\hspace{8pt}}}{12.5cm}
{Variables reclassified from classical to anomalous Cepheids}
{\hline 
\noalign{\vskip3pt}
\multicolumn{1}{c}{Old identifier} & \multicolumn{1}{c}{New identifier} \\
\noalign{\vskip3pt}
\hline
\noalign{\vskip3pt}
OGLE-SMC-CEP-4954 & OGLE-SMC-ACEP-122 \\
OGLE-SMC-CEP-4957 & OGLE-LMC-ACEP-146 \\
OGLE-LMC-CEP-3376 & OGLE-GAL-ACEP-028 \\
\noalign{\vskip3pt}
\hline}

\vspace*{-17pt}
\Section{Collection of Cepheids and RR~Lyr Stars in the Magellanic System}
The complete OGLE collection of variable stars in the Magellanic System
currently consists of 9650 classical Cepheids, 343 type~II Cepheids, 278
anomalous Cepheids (including 10 Galactic variables in front of the
Magellanic System), and 47\,828 RR~Lyr stars. The new variables, their
basic parameters (coordinates, periods, mean brightness, amplitudes,
Fourier coefficients), time-series {\it VI} photometry, and finding charts
have been added to the FTP and WWW sites:
\vskip5pt
\centerline{\it ftp://ftp.astrouw.edu.pl/ogle/ogle4/OCVS/}
\vskip3pt
\centerline{\it http://ogle.astrouw.edu.pl}

We tested the completeness of our collection of classical pulsators by
cross-matching it with the International Variable Star Index (VSX, Watson
\etal 2006) and with the Gaia Data Release~2 catalog of Cepheids and RR~Lyr
stars (Clementini \etal 2019). Currently, the VSX database contains the
most complete compilation of variable stars discovered by various
surveys. In the VSX database we successfully identified 112 variable stars
out of 1403 the newly detected classical pulsators. Most of these stars
were discovered by the Catalina Sky Survey (Drake \etal 2017) and by the
ASAS-SN project (Jayasinghe \etal 2018). We also investigated the OGLE
light curves of stars classified in VSX as Cepheids or RR~Lyr stars which
are absent in the OCVS and we found no new classical pulsators.

\vskip5pt
Gaia Data Release~2 catalog contains 9575 variables classified as
Cepheids (of all types) and 140\,784 as RR~Lyr stars distributed throughout
the sky (Clementini \etal 2019). In the region of the Magellanic Clouds, we
successfully extracted OGLE light curves for 7595 Gaia Cepheids and 34\,046
Gaia RR~Lyr stars, the vast majority of which (7512 Cepheids and 32\,080
RR~Lyr stars) have already been published in the OCVS, although some
confusion between variability types exists between Gaia and OGLE
catalogs. Further three Cepheids and 837 RR~Lyr stars discovered by the
Gaia team are located in the outer fields and are included in this upgrade
of the OCVS.

\vskip5pt
We carefully analyzed the OGLE light curves of the remaining 80 Gaia
candidates for Cepheids and 1129 candidates for RR~Lyr stars which have not
yet been detected in the OGLE data. This analysis has yielded three
additional Cepheids (two classical and one anomalous Cepheid) and 229
RR~Lyr stars in the central regions of the Magellanic Systems. Some of
these variables have been independently noticed by our template-fitting
method. Two of the three previously overlooked Cepheids have pulsation
periods very close to 1~d, which prevented the correct determination of the
period from the ground-based observations due to daily aliasing. The third
missed Cepheid has a very poorly sampled OGLE light curve (only 12 points
in the {\it I}-band), because it is located at the very edge of the OGLE
field. The overlooked RR~Lyr stars have been omitted in the previous
editions of the OCVS because of sparse photometry, or noisy light curves,
or true periods equal to 1/2 or 2/3 of a day. All the positively identified
classical pulsators have been added to the OCVS, yet increasing the high
completeness level of our sample: above 99\% for Cepheids and about 96\%
for RR~Lyr stars.

\vspace*{6pt}
\Section{Discussion}
\vspace*{3pt}
\Subsection{Classical Cepheids}
Classical Cepheids serve as precise standard candles and tracers of the
young stellar population. Recently, the OGLE project published a virtually
complete list of classical Cepheids in the Magellanic System (Soszyñski
\etal 2017) containing 9649 objects. In this investigation, we slightly
modify this list by reclassifying three objects from classical to anomalous
Cepheids and by adding four new detections, two found in the Gaia DR2
catalog and two in the outer fields.

We identified one fundamental-mode and one first-overtone classical Cepheid
in the outer fields, both on the eastern outskirts of the LMC. Fig.~2 shows
the sky distribution of classical Cepheids in the Magellanic System. The
classification of the first-overtone pulsator is somewhat uncertain,
because its light curve is affected by poor phase coverage due to the
period of 0.499~d. In turn, the classification of the fundamental-mode
Cepheid is reliable, taking into account their Fourier coefficient and
positions on the PL diagram. The presence of the young stars so far away
from the LMC center is intriguing, but not unusual. The OCVS contains more
classical Cepheids located at far peripheries of both Magellanic Clouds,
not only in the area of the Magellanic Bridge (Jacyszyn-Dobrzeniecka \etal
2019). The origin of these objects is unknown.
\vskip7pt
\begin{figure}[h]
\begin{center}
\includegraphics[width=12.7cm]{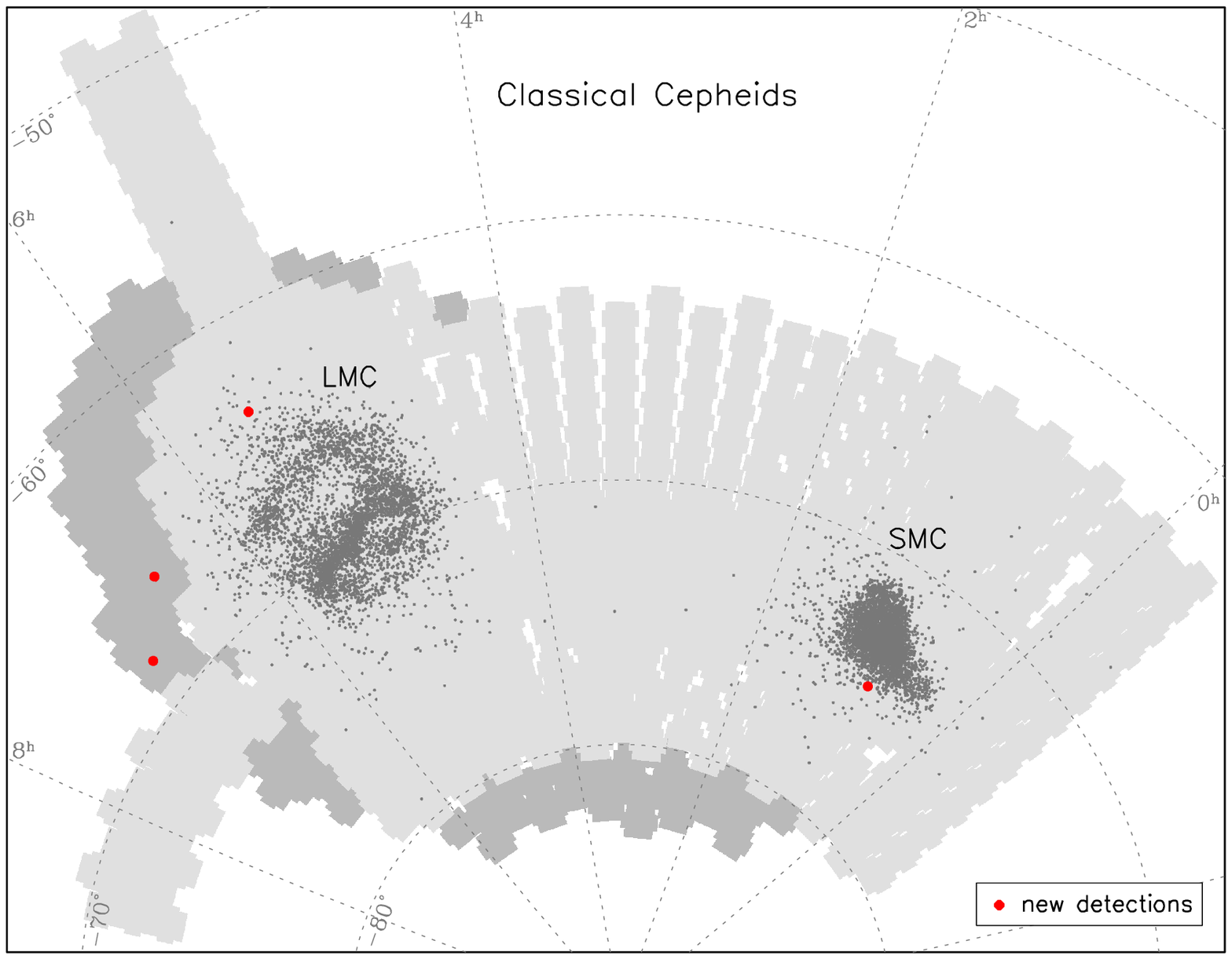}
\end{center}
\vskip3pt
\FigCap{Sky distribution of classical Cepheids in the Magellanic
  System. Gray dots mark objects included in the previous edition of the
  OCVS, while red dots indicate newly detected Cepheids. The gray area
  shows the sky coverage of the OGLE-IV fields (the outer fields are marked
  with a darker shade).}
\end{figure}

\Subsection{Type II Cepheid}
The previous edition of the OCVS contained 340 type~II Cepheids in the
Magellanic System (Soszyñski \etal 2018) categorized to four classes:
BL~Her, W~Vir, peculiar W~Vir, and RV~Tau stars. Recently, the first two
cases of type~II Cepheids (BL~Her stars) pulsating solely in the
first-overtone mode have been discovered in the LMC and added to the OCVS
(Soszyñski \etal 2019). Here, we complement the OCVS with three newly
detected type~II Cepheids in the outer fields, but only two of them are
members of the LMC.

\begin{figure}[b]
\begin{center}
\includegraphics[width=12cm]{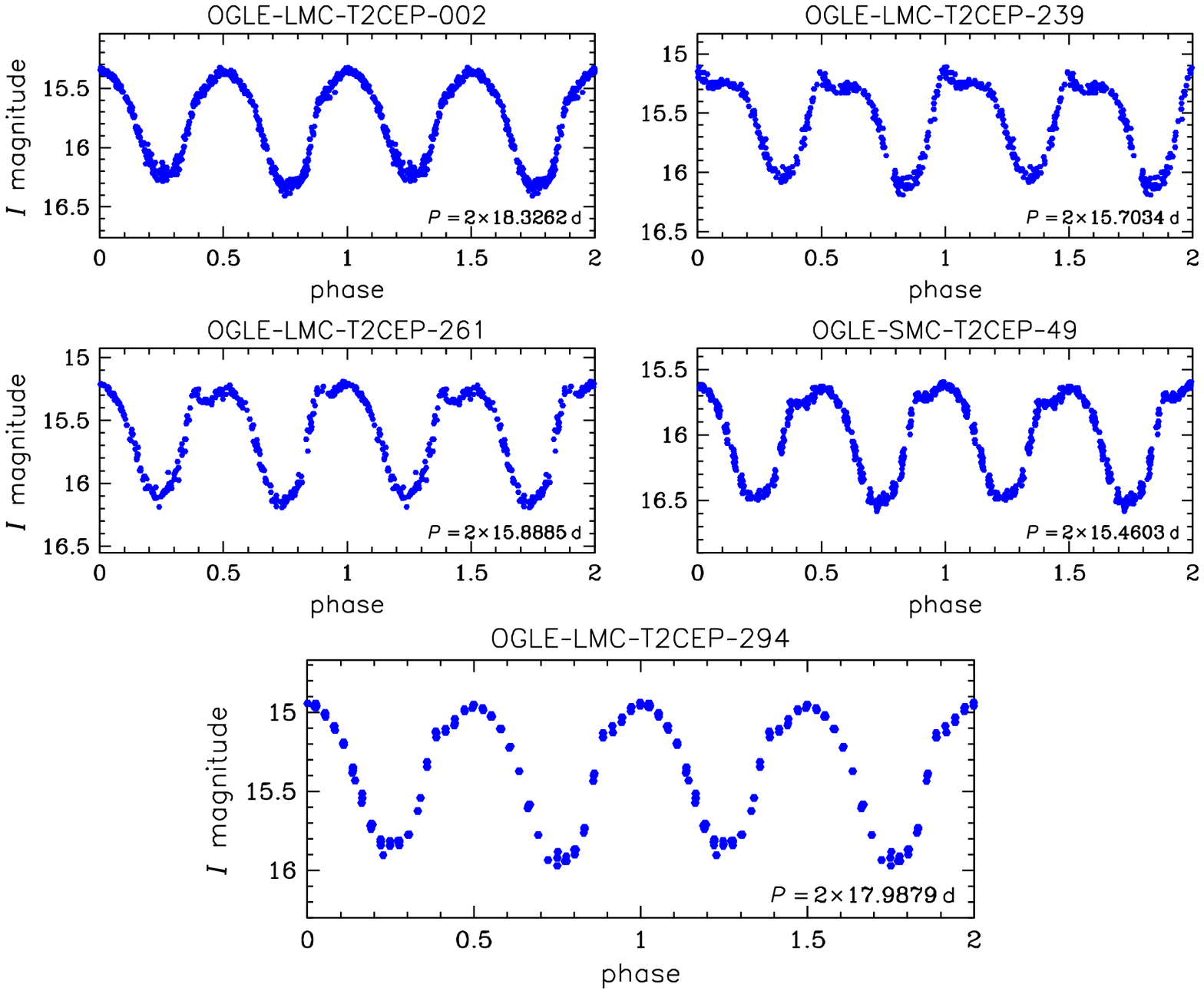}
\end{center}
\vspace*{-11pt}
\FigCap{{\it I}-band light curves of W Vir stars showing alternations of
  deeper and shallower minima (period doubling). {\it Bottom panel} shows the
  light curve of the newly detected OGLE-LMC-T2CEP-294.}
\end{figure}
OGLE-LMC-T2CEP-292 -- a BL~Her star located far southwest of the LMC center
-- is about 1.6~mag brighter than other LMC BL~Her stars with the same
periods. The classification of OGLE-LMC-T2CEP-292 as a BL~Her variable is
rather certain, because its light curve shows characteristic features of
this class. It is unlikely that OGLE-LMC-T2CEP-292 is blended with another
unresolved star, because the amplitude of the light curve ($A_I=0.67$~mag)
is not reduced, as it would be expected from a heavily blended star. The
conclusion is that this BL~Her star belongs to the Milky Way halo and it is
located about 24~kpc from the Sun.

On the eastern side of the LMC, we detected two type~II Cepheids which
probably are associated with this galaxy. OGLE-LMC-T2CEP-293 was
categorized as a peculiar W~Vir star -- a subclass of type~II Cepheids
isolated for the first time by Soszyñski \etal (2008). Again, our
classification was based on the light curve shape. OGLE-LMC-T2CEP-293 is
brighter than ``regular'' W~Vir stars in the LMC, but this is one of the
characteristics of the peculiar W~Vir stars.
\begin{figure}[t]
\begin{center}
\includegraphics[width=12.5cm]{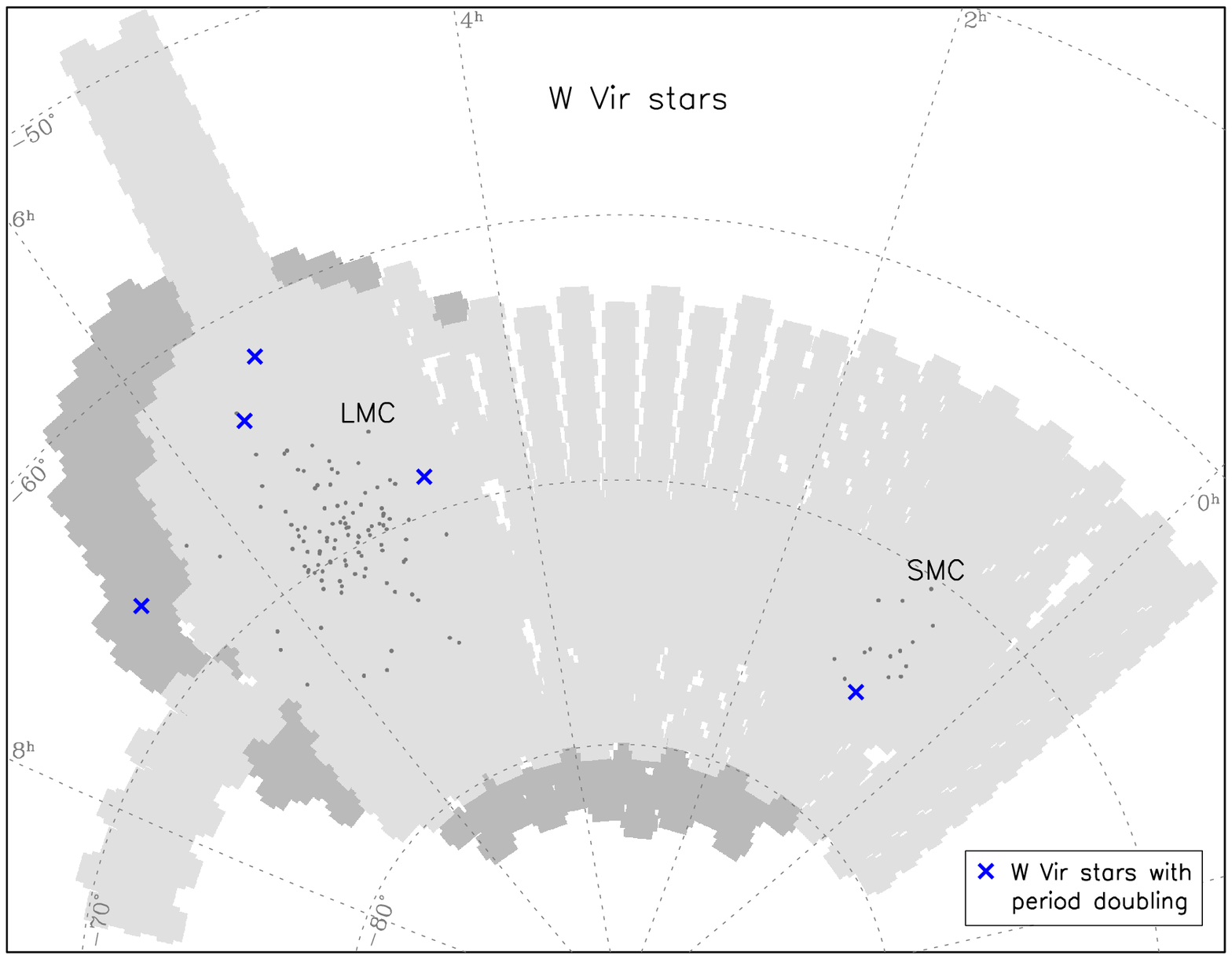}
\end{center}
\FigCap{Sky distribution of W Vir stars in the Magellanic System. Gray dots
  indicate ``regular'' variables (with no period doubling), while blue
  crosses mark W Vir stars showing alternations of deeper and shallower
  minima (the same stars as shown in Fig.~3). The gray area shows the sky
  coverage of the OGLE-IV fields (the outer fields are marked with a darker
  shade).}
\end{figure}

The third new type~II Cepheid in the OCVS -- OGLE-LMC-T2CEP-294 -- fits
well to the PL relation of type~II Cepheids, which proves that it is a
member of the LMC. Since the pulsation period of OGLE-LMC-T2CEP-294 is
shorter than 20~d ($P=17.99$~d), we classify this object as a W~Vir star,
however it belongs to a small group of W~Vir stars exhibiting alternations
of deeper and shallower minima -- a feature typical for RV~Tau stars. There
are several similar W~Vir stars in the Magellanic System (Fig.~3), and most
of them are located rather far from the LMC and SMC centers (Fig.~4), at
the edge of the sky distribution of ``regular'' W~Vir stars. It is worth
noting that W~Vir stars generally are more concentrated toward the centers
of the LMC and SMC than other subclasses of type~II Cepheids (Iwanek \etal
2018). The larger spatial dispersion of period-doubling W~Vir variables may
indicate that they belong to an older stellar population than the
``regular'' W~Vir stars.

\Subsection{Anomalous Cepheid}
OGLE detected the first {\it bona fide} anomalous Cepheids in the
Magellanic Clouds (Soszyñski \etal 2008). The latest edition of the OCVS
(Soszyñski \etal 2017) includes 261\footnote{One object was reclassified by
  Soszyñski \etal (2019) as a first-overtone type II Cepheid.}
fundamental-mode and first-overtone pulsators of that type. Such a large
sample of anomalous Cepheids allowed us to develop a method of
distinguishing between different types of classical pulsators based on
their light curve morphology and to detect the first certain anomalous
Cepheids in the Milky Way halo. So far, seven Galactic anomalous Cepheids
have been discovered in the foreground of the Magellanic Clouds (Soszyñski
\etal 2017) and further 45 in the Galactic bulge and disk (Udalski \etal
2018).

\begin{figure}[htb]
\begin{center}
\includegraphics[width=12.5cm]{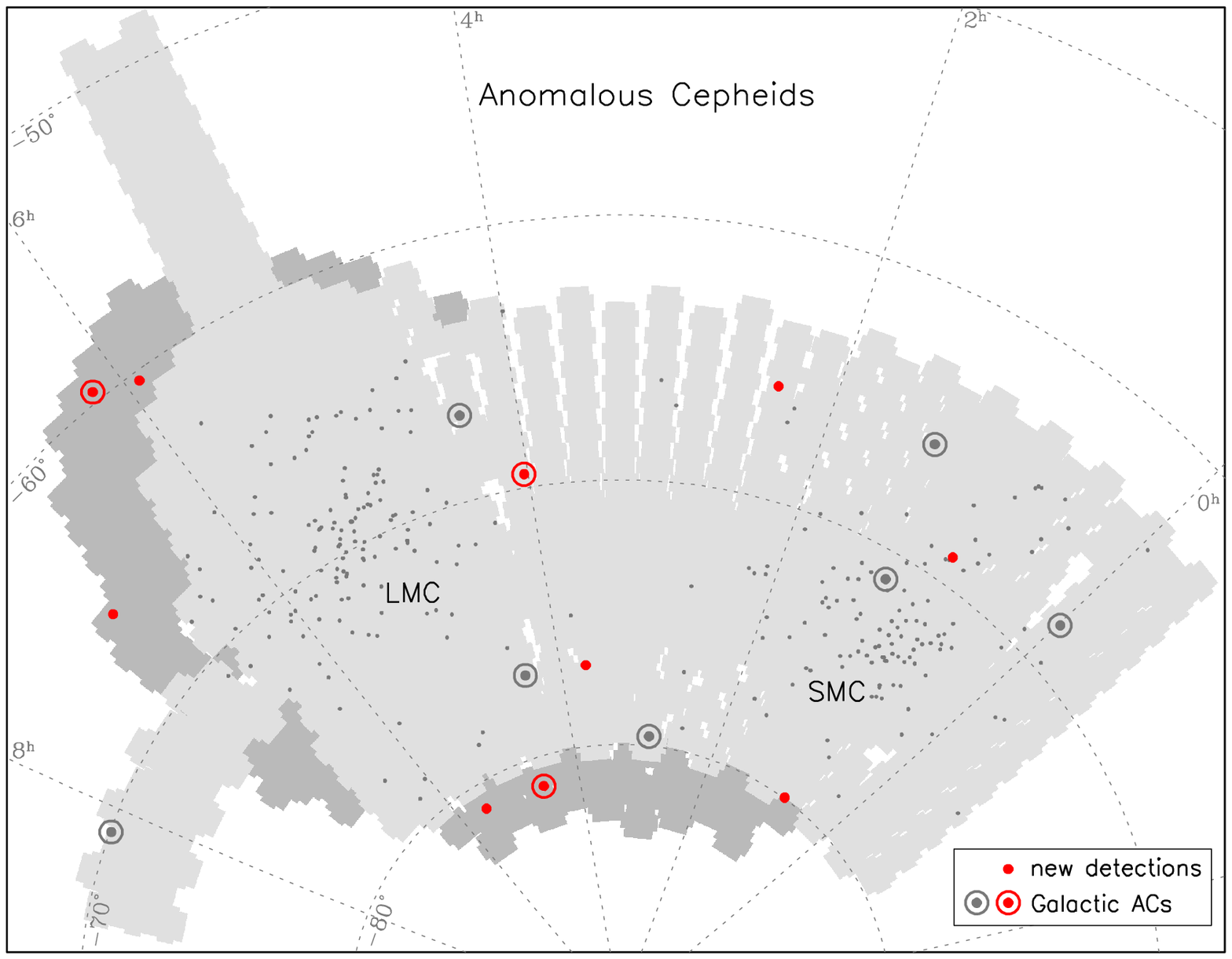}
\end{center}
\FigCap{Sky distribution of anomalous Cepheids in the Magellanic
  System. Gray symbols mark objects included in the previous edition of the
  OCVS, while red symbols indicate newly detected and reclassified
  Cepheids.  Large circles show positions of Galactic anomalous Cepheids in
  the foreground of the Magellanic Clouds. The gray area shows the sky
  coverage of the OGLE-IV fields (the outer fields are marked with a darker
  shade).}
\end{figure}
Here, we supplement the OCVS with 10 new anomalous Cepheids (seven
belonging to the Magellanic Clouds and three to the halo of our Galaxy),
three of which are reclassified classical Cepheids. Six new anomalous
Cepheids have been detected in the outer fields. The sky distribution of
the previously known and new anomalous Cepheids toward the Magellanic
System is shown in Fig.~5. A relatively large number of anomalous Cepheids
at the far peripheries of the Magellanic Clouds proves that they form a
vast halo around both galaxies, so they belong to a very old stellar
population (Iwanek \etal 2018). This in turn indicates that anomalous
Cepheids are remnants of close-binary interactions, probably products of
the coalescence of binary components (\eg Gautschy and Saio 2017).

\begin{figure}[t]
\begin{center}
\includegraphics[width=12.5cm]{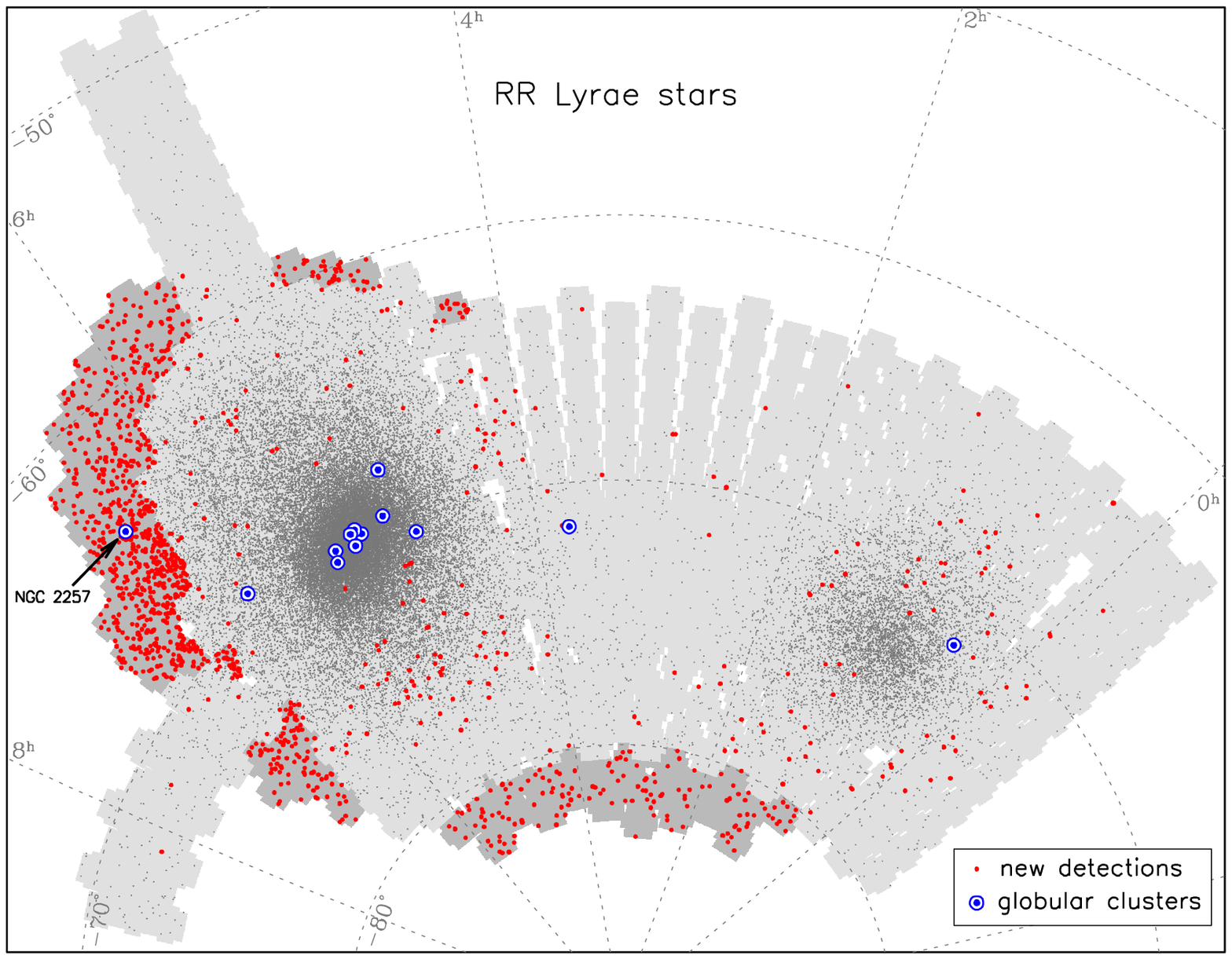}
\end{center}
\FigCap{Sky distribution of RR~Lyr stars in the Magellanic System. Gray
  dots mark objects included in the previous edition of the OCVS, while red
  dots indicate newly detected variables. Blue circles show positions of
  globular clusters hosting RR~Lyr stars. The gray area shows the sky
  coverage of the OGLE-IV fields (the outer fields are marked with a darker
  shade).}
\end{figure}

\Subsection{RR~Lyrae Stars}
RR~Lyr stars -- the tracers of the old stellar population -- are the most
numerous group of classical pulsators. So far, the OCVS (Soszyñski \etal
2016a, 2017) contained in total 46\,442 RR~Lyr variables inside and in
front of the Magellanic System (Fig.~6). With the current upgrade, we
increase this number by 1386 objects (1157 in the outer fields), to
47\,828. 191 of 1386 (14\%) new RR~Lyr stars are brighter than $I=18$~mag
which indicates that they belong to the Milky Way halo.

In the previous edition of the OCVS (Soszyñski \etal 2016a) we listed 12
globular clusters containing RR~Lyr stars. In the current investigation we
extend this list by NGC~2257 -- a globular cluster located on the eastern
side of the LMC. The variable star content of NGC~2257 has been studied for
decades (\eg Alexander 1960, Nemec \etal 1985, 2009). According to the most
recent inventory of variable stars by Nemec \etal (2009), NGC~2257 contains
46 RR~Lyr stars, 23 of which pulsate in the fundamental mode (RRab stars),
20 pulsate in the first-overtone (RRc stars), and three have both modes
excited (RRd stars). We independently detected 44 of these variables. An
RRc star V15 was missed in our search, because it fell into a gap between
CCD chips of the OGLE mosaic camera. Another RRc star, V50, was rejected
because of a very small apparent amplitude ($A_I=0.03$~mag) and red color
($V-I=1.09$~mag). However, Nemec \etal (2009) analyzed the HST photometry
of NGC~2257 and found that V50 is an RRc variable blended by a red giant
star. Thus, we included this object in our collection.

\Section{Conclusions}
In this paper, we provide a final release of the OGLE Collection of
Cepheids and RR~Lyr stars in the Magellanic System. Compared to the
previous edition of the OCVS, the sky coverage has been increased from
about 670 to 765 square degrees by adding 83 OGLE-IV fields in the
outskirts of the Magellanic System. In this region we found 11 Cepheids of
different types and 1157 RR~Lyr stars. Additionally, we supplement the
OCVS with classical pulsators identified by the Gaia project in the central
regions of the Magellanic Clouds. The extended OGLE Collection of classical
pulsators is an indispensable tool for the characterization of the outer
halos of the Magellanic Clouds to trace the history of their interactions
between each other and between Magellanic System and the Milky Way.

\Acknow{We thank Z.~Ko³aczkowski and A.~Schwarzenberg-Czer\-ny for
providing software used in this study.

\hglue-9pt This work has been supported by the National Science Centre, Poland,
grant MA\-ESTRO no. 2016/22/A/ST9/00009. The OGLE project has received
funding from the Polish National Science Centre grant MAESTRO no.
2014/14/A/ST9/00121.}

\end{document}